\documentclass[sigconf]{acmart}
\usepackage{xcolor}
\usepackage{enumitem}
\usepackage{mdframed}
\usepackage{multirow}  
\usepackage{multicol} 
\usepackage{makecell}
\usepackage{caption} 

\AtBeginDocument{%
  \providecommand\BibTeX{{%
    \normalfont B\kern-0.5em{\scshape i\kern-0.25em b}\kern-0.8em\TeX}}}

\setcopyright{acmlicensed}
\copyrightyear{2018}
\acmYear{2018}
\acmDOI{XXXXXXX.XXXXXXX}

\acmConference[Conference acronym 'XX]{conference title}{June 03--05,
  2018}{Woodstock, NY}
\acmISBN{978-1-4503-XXXX-X/18/06}

\begin{document}

% \title{Embedding-based Item Retrieval with Natural Language Interface}
% \title{Embed Any Text for Item Retrieval}
\title{Aligning Language Models for Versatile Text-based Item Retrieval}

\author{Yuxuan Lei}
\email{leiyuxuan@mail.ustc.edu.cn}
\affiliation{%
  \institution{University of Science and Technology of China}
  \streetaddress{Fuxing Road 100}
  \city{Hefei}
  \country{China}
  \postcode{230031}
}

\author{Jianxun Lian}
\email{jianxun.lian@outlook.com}
% \authornote{Corresponding author.}
\affiliation{%
  \institution{Microsoft Research Asia}
  \streetaddress{Danling Street 5}
  \city{Beijing}
  \country{China}
}

\author{Jing Yao} 
\email{jingyao@microsoft.com}
\affiliation{%
  \institution{Microsoft Research Asia}
  \streetaddress{Danling Street 5}
  \city{Beijing}
  \country{China}
}

\author{Mingqi Wu}
\email{mingqi.wu@microsoft.com} 
\affiliation{%
  \institution{Microsoft Gaming} 
  \city{Redmond}
  \country{United States}}

\author{Defu Lian}
\email{liandefu@ustc.edu.cn}
\affiliation{%
  \institution{University of Science and Technology of China}
  \streetaddress{Fuxing Road 100}
  \city{Hefei}
  \country{China}
  \postcode{230031}
}

\author{Xing Xie}
\email{xing.xie@microsoft.com}
\affiliation{%
  \institution{Microsoft Research Asia}
  \streetaddress{Danling Street 5}
  \city{Beijing}
  \country{China}
}

\renewcommand{\shortauthors}{Yuxuan Lei, et al.}

\begin{abstract}
% Item retrieval is a crucial component of recommender systems and search engines. Embedding-based matching is the leading method for this task. The significant advancements in language models have bolstered the capability to embed various forms of text, aiming for a general-purpose representation. 
% This advancement is exemplified by OpenAI's text embedding APIs, which have emerged as a particularly appealing approach due to their effectiveness.
This paper addresses the gap between general-purpose text embeddings and the specific demands of item retrieval tasks. We demonstrate the shortcomings of existing models in capturing the nuances necessary for zero-shot performance on item retrieval tasks. To overcome these limitations, we propose generate in-domain dataset from ten tasks tailored to unlocking models' representation ability for item retrieval. Our empirical studies demonstrate that fine-tuning embedding models on the dataset leads to remarkable improvements in a variety of retrieval tasks. We also illustrate the practical application of our refined model in a conversational setting, where it enhances the capabilities of LLM-based Recommender Agents like Chat-Rec. Our code is available at https://github.com/microsoft/RecAI.
\end{abstract}





\maketitle

\section{Introduction}
\begin{figure}[t]
    \centering
    % \begin{minipage}[t]{0.49\textwidth}
    \includegraphics[width = 0.48\textwidth]{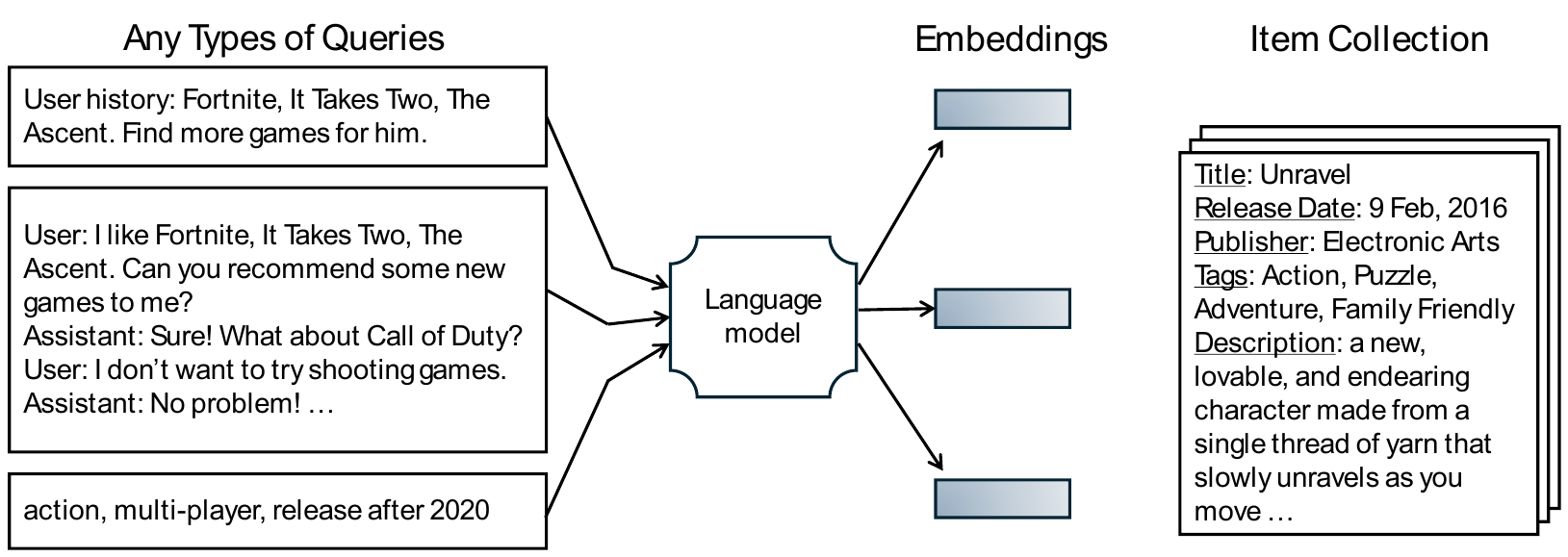}
    % \end{minipage}%
    \vspace{-4ex}
    \caption{Embedding any text for item retrieval.}
    \label{fig:int}
    % \vspace{-2.5ex}
\end{figure}

% Item retrieval is a crucial component of recommender systems and search engines, efficiently selecting a small set of item candidates that align with the current context from an extensive pool of items. The embedding-based matching paradigm, coupled with the two-tower model architecture, has emerged as the predominant approach for item retrieval, gaining popularity with the advancement of deep learning techniques.
Item retrieval is a crucial component of recommender systems and search engines. The embedding-based matching paradigm, coupled with the two-tower model architecture, has emerged as the predominant approach for item retrieval.
Recently, large language models (LLMs) have shown remarkable progress toward artificial general intelligence (AGI).
% , driving innovative approaches to conventional tasks and leading to significant breakthroughs in sophisticated areas such as code intelligence and mathematical problem-solving. 
Owing to the limitations of LLMs, such as a lack of domain-specific data and newly emerged information, there is an increasing need for item retrieval techniques to augment LLMs, in the form of natural text as queries. Here, "items" could be examples for in-context learning, tools that LLM-based agents can leverage, or products offered in a business domain; whereas "queries" might be instructions or conversation context. 
% (1) need to list what is the main challenge in language as query
% (2) compare with ID embedding

Previous studies have attempted to create a general-purpose text embedding model for universal information retrieval. Works like \cite{DBLP:journals/corr/abs-2104-08821} and \cite{neelakantan2022text} employ contrastive pre-training on extensive data to boost the performance of semantic text matching. Similarly, \cite{wang2022text} have put together a large-scale text pair dataset named CCPairs to enhance the quality of data used for contrastive learning. Additionally, \cite{li2023towards} have gathered a supervised fine-tuning dataset that covers a broad range of tasks to improve results. Compared to \cite{li2023towards}, the method by \cite{bge_embedding} includes an extra pre-training step, modeled after RetroMAE~\cite{DBLP:conf/emnlp/XiaoLSC22}, before the contrastive learning phase.

However, general-purpose text embedding models often fall short of achieving satisfactory zero-shot performance for specific tasks, such as item retrieval. Achieving precise performance in item retrieval is critical for online businesses that generate billions in revenue, such as search advertising and recommendation systems. The main reason for their suboptimal performance is that these models tend to produce general semantic representations for text similarity, focusing less on tailoring information to represent relevant items mentioned within the text and disregarding other unrelated details. Furthermore, the query for item retrieval can come in various formats, including detailed descriptions, formatted instructions, or unstructured and sparse attributes. The aim of this paper is to refine a general-purpose text embedding model to enhance item retrieval, regardless of the textual data format, as illustrated in Figure~\ref{fig:int}.

To this end, we propose the creation of a specialized fine-tuning dataset designed for item retrieval, encompassing ten distinct types of tasks. These tasks are designed to capture fundamentally different approaches to representing items, ranging from implicit user behaviors to instructions that convey explicit intents. We carried out experiments using two real-world gaming datasets from Xbox and Steam platforms. The results confirm that after fine-tuning, various language models exhibit significant improvements in item retrieval across a broad spectrum of retrieval tasks. Furthermore, we showcase that the refined model, when deployed for item retrieval within conversational contexts, can effectively facilitate LLM-based Recommender Agents, such as Chat-Rec~\cite{DBLP:journals/corr/abs-2303-14524}.

\section{Methodology}

\subsection{Problem Formulation}
Suppose our database contains $N$ items, denoted as $\mathcal{T}=\{t_1, t_2, ..., t_N\}$. Our objective is to develop an embedding model, $f(\cdot)$, that effectively maps a given query $q$ to the most relevant items within $\mathcal{T}$. The items in $\mathcal{T}$ are thoroughly described by a compilation of their titles, tags, attributes, and descriptions. In contrast, the query $q$ could take various textual forms, ranging from a simple bag of keywords to a complete sentence detailing an item, or even an informal text snippet reflecting a user's interest. Figure~\ref{fig:int} illustrates the task.

\subsection{Fine-tuning Task Collection}
We argue that a general-purpose text embedding model is not optimized for matching items from any types of text, a specialized text embedding model tailored for precise item retrieval is indispensable. Therefore, we propose a set of 10 distinct tasks to form a comprehensive dataset designed to significantly improve a language model's proficiency in item retrieval. Each task is named to reflect the specific type of query it represents.

The first group of tasks consolidate the matching from implicit user preferences to relevant items:

\textbf{User history} (UH2I). The query is represented by the behavior history of a user. E.g., \textsl{A user has played Age of Empires, It Takes Two, and The Ascent. Recommend an item for him to play next}. 

\textbf{Item} (I2I). The query is represented by a source item. E.g., a query can be: \textsl{Games like Call of Duty}.

\textbf{User Summary} (US2I). Instead of directly fill the user history into query, in this task, we let GPT-4 to generate a summary about the user based on his behavior history, then use the GPT-4 summary as query. E.g., \textsl{The user enjoys a diverse range of video games including real-time strategy, cooperative narrative adventures, cyberpunk-themed action RPGs, and competitive battle royale shooters}. 

The second group of tasks match items based on the target item's explicit attributes:

\textbf{Full Attributes} (FA2I). We concatenate the complete details of the item into query, excluding the title and description attributes, and then randomly shuffling the attribute order to enhance robustness. E.g., a query can be: \textsl{What games offer these features? tags : action, pixel art, action, price : 10, release date : march 19, 2021, publisher : prison games, primary genre : classics}.

\textbf{Sparse Attributes} (SA2I). In contrast to Full Attributes, in this task, we randomly sample a few attributes and shuffle them to offer only partial information in the query.

\textbf{Attributes Summary} (AS2I). In the real world, users often describe items using their unique tone and a selection of attributes they consider important. To mimic this behavior, we prompt GPT-4 to craft brief item summaries using a random assortment of attributes for query generation.

The third group of tasks aim to enhance fuzzy queries:

\textbf{Name Misspell} (NM2I). People frequently make typographical errors; therefore, to simulate instances of typos, we randomly add, remove, or replace characters in item names. Additionally, we use GPT-4 to generate several plausible misspellings for each item name to cover more complicate cases.

\textbf{Vague Condition} (VC2I). Users may not always recall precise attributes of items, leading them to employ vague terms when searching. For example, in the query, \textsl{A first-person shooting game released after 2020 with a price under 10 dollars}, the specifications for release date and price are not exact figures. 

\textbf{Negative Attributes} (NA2I). Users may seek items lacking certain features rather than possessing them.  For example, when users specify \textsl{Shooting games but not made for kids},  they are likely seeking intense, competitive games as opposed to family-friendly, cartoon-styled games.

The last task handles queries in which both implicit preferences and explicit intents:

\textbf{User History and Query} (UQ2I). A user's behavior history reflects his general preference, which is a type of implicit signal. Besides, he may have some instant demands, looking for items with specific attributes. Thus, implicit preferences and explicit intents may appear together in a query.
E.g., \textsl{Suggest some local co-operative games for a user who likes It Takes Two, The Ascent, and Fortnite}.

\subsection{Data Generation}
% Item metadata we utilize includes \textsl{title}, \textsl{tags}, \textsl{publisher}, \textsl{developer}, \textsl{genre}, \textsl{price}, \textsl{release date}, and \textsl{description}. 
We employ two datasets for item retrieval: Xbox and Steam. To enhance the diversity of our dataset, we manually create 40 unique prompt templates for each task, 20 for training set and 20 for test set. When constructing the content of a query, we first sample several fields from a given item, and then from these fields, we sample one or more values. This fine-grained sampling approach renders the query more diverse and comprehensive, thereby making the trained model more robust. It is important to note that negatives vary substantially between different tasks. Our preliminary experiments reveal that the widely adopted in-batch negatives strategy often introduces false negatives, which deteriorates the final training outcome. Consequently, we do not employ this strategy and sample 7 true negatives for each data sample. This strategy plays a crucial role in ensuring the stability of the model's performance. We also observe that the model has more difficulty learning tasks closely tied to user behaviors (such as UH2I and I2I) so we generate more data for the them. Overall, on Xbox, the total training data is 120,000, of which the UH2I is 40,000, I2I is 20,000, and the rest are about 7,000 each. On Steam, the total training data is 200,000, of which the UH2I is 80,000, I2I is 40,000, and the rest are about 10,000 each.

% \subsection{Model Selection}

\captionsetup[table]{font={small}}
\begin{table*}[t]
\centering  
\caption{\fontsize{8pt}{8.5pt} Overall performance of different models for various item retrieval tasks, on Xbox and Steam dataset. Cov@5 are abbreviations for Coverage@5. OOD refers to evaluating a model that has been fine-tuned on a domain distinct from the one it is being tested on. We use the "large" version of BERT/ BGE-v1.5/E5 and "7B" version of RepLLaMA.}  
\label{table:overall}  
    \vspace{-3ex}
\setlength{\tabcolsep}{3.5pt} % 设置列间距为10pt  
\fontsize{9.0pt}{10.5pt}\selectfont
\begin{tabular}{c|c|c||c|c|c|c|c||c|c|c|c||c}  
\Xhline{1.5pt}
\multicolumn{3}{c||}{} &  \multicolumn{5}{c||}{Original Models} &  \multicolumn{4}{c||}{Our Finetuned Models} & \multicolumn{1}{c}{\ \ OOD \ \ } \\ \hline
\hline
dataset & task & metric & ada-002 & BERT & BGE-v1.5 & E5   & RepLLaMA  &  BERT & BGE-v1.5 &  E5 
 & RepLLaMA & E5  \\ \hline    
% \multicolumn{12}{c}{Xbox Dataset}   \\ \hline  
% \hline  
\multirow{10}{*}{Xbox} &UH2I & Hit@5  & 0.0818 & 0.0231 & 0.0297 & 0.0853 & 0.0642 & 0.4563 & \textbf{0.4694} & \underline{0.4603} & 0.2905 & 0.0941 \\ \cline{2-13} 
 & I2I  & Hit@5  & 0.2729 & 0.1077 & 0.2661 & 0.2638 & 0.2816 & 0.4878 & \underline{0.5419} &\textbf{0.5477} &0.5013 & 0.3022 \\ \cline{2-13}   
& US2I & Hit@5 & 0.0555 & 0.0005 & 0.0424 & 0.0890 & 0.0565 & 0.4518 & \underline{0.4723} & \textbf{0.4895} & 0.3236&  0.1052\\ \cline{2-13}   
& FA2I & Cov@5 & 0.1553 & 0.0919 & 0.1595 & 0.1760 &0.1728  & 0.1988 & \underline{0.2139} & \textbf{0.2160} & 0.2129& 0.2102\\ \cline{2-13}  
& SA2I & Cov@5 & 0.0893 & 0.0265 & 0.0951 & 0.1443 & 0.1527 & 0.2258 &\underline{0.3022} & \textbf{0.3187}& 0.2985& 0.2755 \\ \cline{2-13}  
& AS2I & Hit@5 &0.3562  & 0.0856 &  0.3772& 0.4065 & 0.3988 & 0.4281 &\underline{0.4800} &\textbf{0.5027} & 0.4734& 0.4674 \\ \cline{2-13}   
& NM2I & Hit@5 & 0.9605 & 0.0381 & 0.8027 & 0.8430 &\underline{0.9587}  &  0.7970&0.8835 & 0.9111& \textbf{0.9686}& 0.9205\\ \cline{2-13} 
& VC2I & Cov@5 & 0.5138 & 0.4966 & 0.5158 & 0.5299 & 0.5366 & 0.8467 & \underline{0.9769}& \textbf{0.9813}& 0.9630&0.7701 \\ \cline{2-13} 
& NA2I & Cov@5 & 0.5014 & 0.7305 & 0.5253 & 0.5450 &0.4039  & 0.9598 & \textbf{0.9912}&\underline{0.9778} & 0.9478& 0.8855\\ \cline{2-13}  
& UQ2I & Hit@5 & 0.3807 & 0.0674 & 0.3556 & 0.4788 & 0.3711 & 0.8655 & \underline{0.8984} &\textbf{0.9115} &0.8657 &0.6732 \\ \hline  
\hline  
% \multicolumn{12}{c}{Steam Dataset}   \\ \hline  
% \hline  
dataset & task & metric & ada-002 & BERT & BGE-v1.5 & E5   & RepLLaMA  &  BERT & BGE-v1.5 &  E5 
 & RepLLaMA & E5  \\ \hline  
\multirow{10}{*}{Steam} & UH2I & Hit@5  & 0.0192 & 0.0014 & 0.0179 & 0.0170 & 0.0146 & 0.0405 & \textbf{0.0461} & \underline{0.0457} & 0.0454 & 0.0246 \\ \cline{2-13}  
& I2I  & Hit@5  & 0.2789 & 0.0955 & 0.2828 & 0.2925 & 0.2859 & 0.3963 & \underline{0.4387} & \textbf{0.4583}& 0.3720& 0.2819 \\ \cline{2-13}  
& US2I & Hit@5 & 0.0124 & 0.0000 & 0.0155 & 0.0140 & 0.0150 & 0.0388 & \textbf{0.0466}&\underline{0.0440} &0.0378 & 0.0212 \\ \cline{2-13} 
& FA2I & Cov@5 & 0.1643 & 0.0923 & 0.1776 & 0.1896 & 0.1793 &  0.2294& \underline{0.2430} & \textbf{0.2484}&0.2388 & 0.2260\\ \cline{2-13}
& SA2I & Cov@5 & 0.1246 & 0.0408 &0.1514  &0.1949  &0.1714  & 0.3256 & \underline{0.3731}& \textbf{0.3921}& 0.3518& 0.3133 \\ \cline{2-13} 
& AS2I & Hit@5 & 0.3080 & 0.0468 & 0.3153 & 0.3358 & 0.3385 &0.3509  &0.4286 & \textbf{0.4627}&\underline{0.4318} & 0.3822 \\ \cline{2-13}
& NM2I & Hit@5 & 0.9231 &  0.0260& 0.7567 & 0.8067 & \underline{0.9393} & 0.7717 & 0.8344& 0.8856&\textbf{0.9414} & 0.8623\\ \cline{2-13}
& VC2I & Cov@5 & 0.6046 &0.5470  & 0.5739 & 0.5759 &0.6238  & 0.9819 &\textbf{0.9857} & \underline{0.9828}&0.9635 &0.9368 \\ \cline{2-13}
& NA2I & Cov@5 &0.4381  &0.6826  & 0.4372 & 0.4297 & 0.3179 & \textbf{0.9719} & 0.9611&0.9205 &\underline{0.9658} &0.9573 \\ \cline{2-13}
& UQ2I & Hit@5 &0.1849  & 0.0284 &0.2253  & 0.2648 &0.1661  &0.5699  & \underline{0.6293} &\textbf{0.6627} &0.5577 &0.5265 \\ \Xhline{1.5pt} 
\end{tabular}  
\end{table*} 

\section{experiments}

\subsection{Settings}
\textbf{Baselines and Metrics.} We select several representative models for experimentation. From closed-source models, we include OpenAI text-embedding-ada-002. For open-source models, we employ BERT~\cite{DBLP:conf/naacl/DevlinCLT19} (not specifically designed for text embedding), BGE-v1.5~\cite{bge_embedding} and E5~\cite{wang2022text} (LMs for general text embedding), as well as RepLLaMA~\cite{ma2023fine} (an LLM for general text embedding). We use top-5 hit ratio (Hit@5) and top-5 coverage (Coverage@5) for evaluation. The meaning of Coverage@K is the proportion of items that fully meet the query conditions among the top-K items.
\\
\textbf{Implementation details.} we set the max query length to 512 tokens and the max item length to 256 and 128 for Xbox and Steam dataset respectively. The learning rate is 3e-5 with a warm-up ratio of 10\%. The models are trained for 3 epochs with an overall batch size of 64. Notably, for training the LLMs (RepLLaMA), we employ LoRA and Flash Attention 2 techniques to handle the large computational demands of the LLMs.

\subsection{Overall Results}
Table~\ref{table:overall} reports the overall performance of different models on the Xbox and Steam dataset. We have the following observations:
\begin{itemize}[leftmargin=*]
    \item In-domain fine-tuning is essential for enhancing item retrieval performance. We apply the same fine-tuning technique to four distinct backbone models, each showing substantial improvements over their respective original versions, which operated in a zero-shot setting. For instance, the Hit@5 metric for E5 on the US2I task increased dramatically from 0.0424 to 0.4723, marking a significant qualitative leap. Furthermore, the enhancements achieved through fine-tuning are consistent across all evaluated tasks, suggesting that the refined model acts as a robust and versatile backbone for various item retrieval tasks.
    \item Among the enhanced models, E5 and BGE-v1.5 exhibit superior overall performance compared to BERT and RepLLaMA, which is to be expected. E5 and BGE-v1.5 incorporate a large-scale contrastive learning phase during pretraining, significantly aiding the development of their language model representations. In contrast, BERT's pretraining does not involve contrastive learning, and RepLLaMA is limited to a contrastive pretraining size of only 500k examples. Additionally, the quantity of negative samples plays a crucial role in the efficacy of contrastive fine-tuning. However, due to RepLLaMA's model size, we can only utilize one negative sample per positive instance, which severely restricts its performance potential.
    \item 
    % Among the compared models, only BERT is not pretrained for learning embeddings. Consequently, among the results of original models, BERT exhibits significantly lower performance on almost all tasks when compared with the other four models, with the notable exception of the NA2I task.  Interestingly, the NA2I task involves the inclusion of negative terms within the text, such as \textsl{Shooting games but not made for kids}. Current models struggle to grasp the true semantic intent of such phrases, often placing undue emphasis on keywords like \textsl{Shooting games} and \textsl{made for kids}. However, BERT, which in our experiments employs a simple mean pooling over all tokens, manages to surprisingly retain the meaning of the word \textsl{not}, showing a potential advantage in contexts requiring nuanced interpretation of negations.
    Among the models compared, BERT, lacking pretraining for embedding learning, generally underperforms other original models across tasks, except for the NA2I task which involve negative terms. Original models have difficulty accurately interpreting phrases with negations, tending to focus on keywords and missing the deep semantic intent. In contrast, BERT, using a straightforward mean pooling approach in our settings, effectively captures the significance of "not", indicating a potential strength in processing nuanced negation.
\end{itemize}

\subsection{Out-Of-Domain Test}
% Another critical research question is the extent to which a model fine-tuned on one dataset can adapt to another, distinctly different dataset. This scenario constitutes an Out-of-Domain (OOD) test designed to assess the model's generalization capabilities. The "OOD" columns in Tables~\ref{table:overall}  presents the results of these experiments, where, for instance, OOD results in the rows related to Xbox indicate that the model was fine-tuned on the Steam dataset and subsequently tested on the Xbox dataset. The results lead two notable insights. First, for tasks closely tied to user behaviors, such as UH2I and I2I, models tested OOD exhibit weak performance. For instance, in the Xbox domain, the raw E5 model scores 0.0853 on the UH2I task. While an E5 model fine-tuned on the Steam domain shows a marginal improvement with a score of 0.0941, it still lags considerably behind an E5 model fine-tuned within the Xbox domain, which scores 0.4603. This suggests that user behaviors are highly domain-specific and do not easily transfer across domains based on generalized assumptions. Second, for tasks that are less influenced by user behavior, such as SA2I and NA2I, the fine-tuned models exhibit a robust capacity for generalization. This finding encourages the integration of such tasks into the foundational pretraining of models, as they seem to capture universal logic that transcends specific domain contexts.
We investigate the adaptability of a model fine-tuned on one dataset when applied to a different dataset, an Out-of-Domain (OOD) test of generalization. The "OOD" columns in Tables~\ref{table:overall} show these experiments' results, such as a model fine-tuned on Steam and directly tested on Xbox. Two key insights emerge: First, models exhibit poor OOD performance on tasks closely related to user behaviors, like UH2I and I2I. For instance, the E5 model fine-tuned on Steam shows only a slight improvement over the original E5 in the Xbox UH2I task and remains significantly inferior to the E5 model fine-tuned specifically for Xbox, indicating that user behaviors are domain-specific and resist generalization. Second, models demonstrate strong generalization on tasks less dependent on user behavior, such as SA2I and NA2I. This robustness supports incorporating such tasks into foundational pretraining, as they appear to capture a universal logic applicable across domains.

\subsection{Results with Recommender AI Agent}
To verify whether the finetuned model can truely embed any text for item retrieval, we assessed its performance in a practical application: a Recommender AI Agent. The agent is tasked with recommending items based on a user's conversation history. We utilized GPT-4 to simulate users, and use the Chat-Rec framework~\cite{DBLP:journals/corr/abs-2303-14524} to function as the recommender agent. The embedding models are required to retrieve relevant items using raw conversational text as input. For evaluation protocol, we adopted the one-turn setting outlined in \cite{huang2023recommender}. The results, detailed in Table~\ref{table:Recommender-Agent}, confirm our expectations: the fine-tuned E5 model substantially surpasses both the original E5 and OpenAI's text-ada-002 in performance.

\begin{table}[t]
\centering  
\caption{Evaluation with Chat-Rec. The metric is Hit@5.}  
    \vspace{-3ex}
\label{table:Recommender-Agent}  
% \setlength{\tabcolsep}{4.5pt} % 设置列间距为10pt  
% \fontsize{10.5pt}{12.5pt}\selectfont
\begin{tabular}{|c|c|c|c|}  
\hline  
  & text-ada-002 & E5$_{large}$ & Finetuned E5$_{large}$ \\ \hline  
Xbox Domain & 0.308 &  0.362  & \textbf{0.884} \\ \hline  
Steam  Domain & 0.198  &  0.268 & \textbf{0.702} \\ \hline  
\end{tabular}  
\end{table} 

\subsection{Case Studies}
We present case studies of the E5 model before and after fine-tuning in Table~\ref{table:query_case}, highlighting two composite instructions. The pre-tuned model often misses the query's full scope and misinterprets negative and ambiguous terms, resulting in incorrect responses. In contrast, the fine-tuned E5 model accurately comprehends all requirements, even for composite tasks like FA2I+NA2I not included in its training, showcasing its excellent generalization ability.
% In the Table~\ref{table:conversation_case}, we also showcases an example from Recommender AI Agent scenario. Similar to the previous findings, the fine-tuned E5 is able to more accurately mine the user's requirements from the dialogue and return items that meet these requirements. However, the current model struggles to extract information from the conversation that would be useful for the item retrieval task.
Additionally, Table~\ref{table:conversation_case}, demonstrates that the fine-tuned E5 model more effectively discerns user needs from dialogue in a Recommender AI Agent scenario, unlike the original model, which struggles to gather relevant information for item retrieval.

% \begin{figure}[htb]
%     \centering
%     \includegraphics[width=0.9\columnwidth]{fig/conv.pdf}
%     % \vspace{-1.5ex}
%     \caption{Case study for conversations.}
%     \label{fig:conv}
% \end{figure}

% \begin{figure}[htb]
%     \centering
%     \includegraphics[width=0.9\columnwidth]{fig/other_case.pdf}
%     % \vspace{-1.5ex}
%     \caption{Case study for conversations.}
%     \label{fig:other_case}
% \end{figure} 

\begin{table}[htbp] 
\centering  
% \captionsetup{font={fontsize=8pt,selectfont}}  
\caption{ Two complex queries, including FA2I+NA2I and FA2I+VC2I. Green text indicates item attributes meet the query requirements, red text indicates non-compliance.} 
    \vspace{-3ex}
\label{table:query_case}  
\setlength{\tabcolsep}{2.0pt} % 设置列间距为10pt 
\fontsize{8.0pt}{8.5pt}\selectfont
\begin{tabular}{l|l}  
\toprule
% \textbf{Query} & \makecell[l]{Search for games with exploration and science fiction tags, \\developed by M2.}
% \\ \midrule
% \textcolor{blue}{\textbf{Ours}} & \makecell[l]{BATTLE GAREGGA Rev.2016 \\ (\textcolor[RGB]{18,200,170}{tags: exploration, science fiction,...; developer: M2})}

%  \\ \midrule
%  \textcolor{blue}{\textbf{E5}} &\makecell[l]{ Curious Expedition \\ (\textcolor{red}{no science fiction elements; not developed by M2})}
% \\ \midrule
\textbf{Query} & \makecell[l]{I'd like to find some shooting games that are not made for kids \\and not 2D platformers.
}
\\ \midrule
\textcolor{blue}{\textbf{Ours}} & \makecell[l]{
Splitgate \\ (\textcolor[RGB]{18,200,170}{genre: shooter; tags: 3D; Not Made for Kids})}

 \\ \midrule
\textcolor{blue}{\textbf{E5}} & NOT A HERO SUPER SNAZZY EDITION (\textcolor{red}{2D game})
\\ \midrule
 \textbf{Query} & \makecell[l]{I'm looking for a sports game, with high quality graphics and  \\ soundtrack, released after 2021.}
\\ \midrule
\textcolor{blue}{\textbf{Ours}} & \makecell[l]{Tour de France 2023 \\(\textcolor[RGB]{18,200,170}{tags: high quality soundtrack, sports; release date: June 08, 2023})}

 \\ \midrule
\textcolor{blue}{\textbf{E5}} & Super Sports Blast (\textcolor{red}{released in 2020})

 \\ \bottomrule 
\end{tabular}  
\end{table}

\begin{table}[htbp]
\centering  
\caption{An conversation of Recommender AI Agent.} 
    \vspace{-3ex}
\label{table:conversation_case}  
\setlength{\tabcolsep}{2.0pt} % 设置列间距为10pt 
\fontsize{8.0pt}{8.5pt}\selectfont
\begin{tabular}{ll}  
\toprule
\multicolumn{2}{c}{Query Context}
\\ \midrule
\textbf{User:}  & \makecell[l]{Hey, I am looking for a recommendation for a new game.  \\I've played Minecraft, Fortnite, Rocket League, and ARK \\Survival Evolved in the past.}  \\ \midrule
\textbf{Assist:} &\makecell[l]{Sure! What type of game or features are you looking for in \\ your next game? }
 \\ \midrule
\textbf{User:} &\makecell[l]{I'm looking for something with 3D graphics, high quality \\soundtrack, and both men's and women's soccer gameplay.}
 \\ \midrule
\textbf{Assist:} & \makecell[l]{Great, do you prefer a singleplayer or multiplayer experience, \\or perhaps both?}
 \\ \midrule
\textbf{User:}& \makecell[l]{I'd like both singleplayer and multiplayer options, including \\local and online competitive play.}
\\ \midrule 
\multicolumn{2}{c}{Response}
\\ \midrule
\textcolor{blue}{\textbf{Ours:}}& EA SPORTS FIFA 23 Xbox One (\textcolor[RGB]{18,200,170}{target item})
\\ \midrule  
 \textcolor{blue}{\textbf{E5:}}& Emily Wants To Play (\textcolor{red}{not sports game; no multiplayer option})
 \\ \bottomrule   
\end{tabular}  
\end{table}

\section{Conclusion}
In conclusion, our study identifies a discrepancy between general-purpose text embeddings and the nuanced representations required for effective item retrieval. To bridge this gap, we have introduced a diverse set of in-domain fine-tuning tasks that elevate language models' proficiency in generating representations tailored for item retrieval. Experimental validation on two real-world datasets confirms that our approach not only enhances retrieval performance but does so using a single, unified model. This marks a substantial advancement in the evolution of more advanced and effective search and recommender systems, as well as in the realm of human-like interactive platforms like Chat-Rec.

\bibliographystyle{ACM-Reference-Format}
\bibliography{myref}


\begin{thebibliography}{10}

%%% the Web address.

\ifx \showCODEN    \undefined \def \showCODEN     #1{\unskip}     \fi
\ifx \showDOI      \undefined \def \showDOI       #1{#1}\fi
\ifx \showISBNx    \undefined \def \showISBNx     #1{\unskip}     \fi
\ifx \showISBNxiii \undefined \def \showISBNxiii  #1{\unskip}     \fi
\ifx \showISSN     \undefined \def \showISSN      #1{\unskip}     \fi
\ifx \showLCCN     \undefined \def \showLCCN      #1{\unskip}     \fi
\ifx \shownote     \undefined \def \shownote      #1{#1}          \fi
\ifx \showarticletitle \undefined \def \showarticletitle #1{#1}   \fi
\ifx \showURL      \undefined \def \showURL       {\relax}        \fi
% invisible to TeX
\providecommand\bibfield[2]{#2}
\providecommand\bibinfo[2]{#2}
\providecommand\natexlab[1]{#1}
\providecommand\showeprint[2][]{arXiv:#2}

\bibitem[Devlin et~al\mbox{.}({[n.\,d.]})]%
        {DBLP:conf/naacl/DevlinCLT19}
\bibfield{author}{\bibinfo{person}{Jacob Devlin}, \bibinfo{person}{Ming{-}Wei Chang}, \bibinfo{person}{Kenton Lee}, {and} \bibinfo{person}{Kristina Toutanova}.} \bibinfo{year}{[n.\,d.]}\natexlab{}.
\newblock \showarticletitle{{BERT:} Pre-training of Deep Bidirectional Transformers for Language Understanding}. In \bibinfo{booktitle}{\emph{NAACL-HLT 2019, Minneapolis, MN, USA, June 2-7, 2019.}}
\newblock


\bibitem[Gao et~al\mbox{.}(2021)]%
        {DBLP:journals/corr/abs-2104-08821}
\bibfield{author}{\bibinfo{person}{Tianyu Gao}, \bibinfo{person}{Xingcheng Yao}, {and} \bibinfo{person}{Danqi Chen}.} \bibinfo{year}{2021}\natexlab{}.
\newblock \showarticletitle{SimCSE: Simple Contrastive Learning of Sentence Embeddings}.
\newblock \bibinfo{journal}{\emph{CoRR}}  \bibinfo{volume}{abs/2104.08821} (\bibinfo{year}{2021}).
\newblock
\showeprint[arXiv]{2104.08821}
\urldef\tempurl%
\url{https://arxiv.org/abs/2104.08821}
\showURL{%
\tempurl}


\bibitem[Gao et~al\mbox{.}(2023)]%
        {DBLP:journals/corr/abs-2303-14524}
\bibfield{author}{\bibinfo{person}{Yunfan Gao}, \bibinfo{person}{Tao Sheng}, \bibinfo{person}{Youlin Xiang}, \bibinfo{person}{Yun Xiong}, \bibinfo{person}{Haofen Wang}, {and} \bibinfo{person}{Jiawei Zhang}.} \bibinfo{year}{2023}\natexlab{}.
\newblock \showarticletitle{Chat-REC: Towards Interactive and Explainable LLMs-Augmented Recommender System}.
\newblock \bibinfo{journal}{\emph{CoRR}}  \bibinfo{volume}{abs/2303.14524} (\bibinfo{year}{2023}).
\newblock
\urldef\tempurl%
\url{https://doi.org/10.48550/ARXIV.2303.14524}
\showDOI{\tempurl}
\showeprint[arXiv]{2303.14524}


\bibitem[Huang et~al\mbox{.}(2023)]%
        {huang2023recommender}
\bibfield{author}{\bibinfo{person}{Xu Huang}, \bibinfo{person}{Jianxun Lian}, \bibinfo{person}{Yuxuan Lei}, \bibinfo{person}{Jing Yao}, \bibinfo{person}{Defu Lian}, {and} \bibinfo{person}{Xing Xie}.} \bibinfo{year}{2023}\natexlab{}.
\newblock \showarticletitle{Recommender ai agent: Integrating large language models for interactive recommendations}.
\newblock \bibinfo{journal}{\emph{arXiv preprint arXiv:2308.16505}} (\bibinfo{year}{2023}).
\newblock


\bibitem[Li et~al\mbox{.}(2023)]%
        {li2023towards}
\bibfield{author}{\bibinfo{person}{Zehan Li}, \bibinfo{person}{Xin Zhang}, \bibinfo{person}{Yanzhao Zhang}, \bibinfo{person}{Dingkun Long}, \bibinfo{person}{Pengjun Xie}, {and} \bibinfo{person}{Meishan Zhang}.} \bibinfo{year}{2023}\natexlab{}.
\newblock \showarticletitle{Towards general text embeddings with multi-stage contrastive learning}.
\newblock \bibinfo{journal}{\emph{arXiv preprint arXiv:2308.03281}} (\bibinfo{year}{2023}).
\newblock


\bibitem[Ma et~al\mbox{.}(2023)]%
        {ma2023fine}
\bibfield{author}{\bibinfo{person}{Xueguang Ma}, \bibinfo{person}{Liang Wang}, \bibinfo{person}{Nan Yang}, \bibinfo{person}{Furu Wei}, {and} \bibinfo{person}{Jimmy Lin}.} \bibinfo{year}{2023}\natexlab{}.
\newblock \showarticletitle{Fine-tuning llama for multi-stage text retrieval}.
\newblock \bibinfo{journal}{\emph{arXiv preprint arXiv:2310.08319}} (\bibinfo{year}{2023}).
\newblock


\bibitem[Neelakantan et~al\mbox{.}(2022)]%
        {neelakantan2022text}
\bibfield{author}{\bibinfo{person}{Arvind Neelakantan}, \bibinfo{person}{Tao Xu}, \bibinfo{person}{Raul Puri}, \bibinfo{person}{Alec Radford}, \bibinfo{person}{Jesse~Michael Han}, \bibinfo{person}{Jerry Tworek}, \bibinfo{person}{Qiming Yuan}, \bibinfo{person}{Nikolas Tezak}, \bibinfo{person}{Jong~Wook Kim}, \bibinfo{person}{Chris Hallacy}, {et~al\mbox{.}}} \bibinfo{year}{2022}\natexlab{}.
\newblock \showarticletitle{Text and code embeddings by contrastive pre-training}.
\newblock \bibinfo{journal}{\emph{arXiv preprint arXiv:2201.10005}} (\bibinfo{year}{2022}).
\newblock


\bibitem[Wang et~al\mbox{.}(2022)]%
        {wang2022text}
\bibfield{author}{\bibinfo{person}{Liang Wang}, \bibinfo{person}{Nan Yang}, \bibinfo{person}{Xiaolong Huang}, \bibinfo{person}{Binxing Jiao}, \bibinfo{person}{Linjun Yang}, \bibinfo{person}{Daxin Jiang}, \bibinfo{person}{Rangan Majumder}, {and} \bibinfo{person}{Furu Wei}.} \bibinfo{year}{2022}\natexlab{}.
\newblock \showarticletitle{Text embeddings by weakly-supervised contrastive pre-training}.
\newblock \bibinfo{journal}{\emph{arXiv preprint arXiv:2212.03533}} (\bibinfo{year}{2022}).
\newblock


\bibitem[Xiao et~al\mbox{.}(2022)]%
        {DBLP:conf/emnlp/XiaoLSC22}
\bibfield{author}{\bibinfo{person}{Shitao Xiao}, \bibinfo{person}{Zheng Liu}, \bibinfo{person}{Yingxia Shao}, {and} \bibinfo{person}{Zhao Cao}.} \bibinfo{year}{2022}\natexlab{}.
\newblock \showarticletitle{RetroMAE: Pre-Training Retrieval-oriented Language Models Via Masked Auto-Encoder}. In \bibinfo{booktitle}{\emph{EMNLP 2022, Abu Dhabi, United Arab Emirates.}} \bibinfo{publisher}{Association for Computational Linguistics}, \bibinfo{pages}{538--548}.
\newblock


\bibitem[Xiao et~al\mbox{.}(2023)]%
        {bge_embedding}
\bibfield{author}{\bibinfo{person}{Shitao Xiao}, \bibinfo{person}{Zheng Liu}, \bibinfo{person}{Peitian Zhang}, {and} \bibinfo{person}{Niklas Muennighoff}.} \bibinfo{year}{2023}\natexlab{}.
\newblock \bibinfo{title}{C-Pack: Packaged Resources To Advance General Chinese Embedding}.
\newblock
\newblock
\showeprint[arxiv]{2309.07597}~[cs.CL]


\end{thebibliography}

\end{document}